\documentclass[aps,prx,twocolumn,10pt]{revtex4-2}
\usepackage{ragged2e}
\usepackage{xcolor}
\usepackage{babel}
\usepackage{amsmath}
\usepackage{amssymb}
\usepackage{stmaryrd}
\usepackage[pdftex]{graphicx}
\usepackage[utf8x]{inputenc}
\usepackage{wasysym}
\usepackage{bbold}
\usepackage{mathtools}
\usepackage{multirow}
\usepackage{nicematrix}
\usepackage{hhline}
\usepackage{blkarray, bigstrut}
\usepackage{float}
\usepackage[colorlinks=true]{hyperref}

\ifpdf
  \pdftrailerid{}
  \hypersetup{
    pdfcreator={},
    pdfproducer={}
  }
\fi

\definecolor{pyplotc0}{rgb}{0.122,0.467,0.706}
\definecolor{pyplotc1}{rgb}{1.000,0.498,0.055}
\definecolor{pyplotc2}{rgb}{0.173,0.627,0.173}
\definecolor{pyplotc3}{rgb}{0.839,0.153,0.157}
\definecolor{pyplotc4}{rgb}{0.580,0.404,0.741}
\definecolor{pyplotc5}{rgb}{0.549,0.337,0.294}
\definecolor{pyplotc6}{rgb}{0.890,0.467,0.761}
\definecolor{pyplotc7}{rgb}{0.498,0.498,0.498}
\definecolor{pyplotc8}{rgb}{0.737,0.741,0.133}
\definecolor{pyplotc9}{rgb}{0.090,0.745,0.812}

\newif\ifhighlight
\highlightfalse

\ifhighlight

\else

\fi
\newcommand{\Duke}{Duke Quantum Center, Department of Electrical and Computer Engineering and Department of Physics \\ Duke University, Durham, NC 27708}
\newcommand{\Yb}{$^{171}\textrm{Yb}^+$}

\begin{document}

\graphicspath{{./imgs/}}

\title{In-situ mid-circuit qubit measurement and reset in a \\ single-species trapped-ion quantum computing system}
\date{\today}

\author{Yichao~Yu}
\email{Corresponding author: yichao.yu@duke.edu}
\affiliation{\Duke}
\author{Keqin~Yan}
\affiliation{\Duke}
\author{Debopriyo~Biswas}
\affiliation{\Duke}
\author{Vivian~Ni~Zhang}
\affiliation{\Duke}
\author{Bahaa~Harraz}
\affiliation{\Duke}
\author{Crystal~Noel}
\affiliation{\Duke}
\author{Christopher~Monroe}
\affiliation{\Duke}
\author{Alexander~Kozhanov}
\affiliation{\Duke}

\begin{abstract}
  We implement in-situ mid-circuit measurement and reset (MCMR) operations on a full-scale trapped-ion quantum computing system by using metastable qubit states in \Yb ions.
  We compare two methods for isolating data qubits from measured qubits: one shelves the data qubits into the metastable state and the other drives the measured qubit to the metastable state without disturbing the other qubits.
  We experimentally demonstrate both methods on a crystal of two \Yb ions using both the $S_{1/2}$ ground state hyperfine clock qubit and the $S_{1/2}$-$D_{3/2}$ optical qubit.
  These MCMR methods result in errors on the data qubit of about $2\%$ without degrading the measurement fidelity.
  With straightforward reductions in laser noise, these errors can be suppressed to less than $0.1\%$.
  The demonstrated methods allow MCMR to be performed in a single-species ion chain without shuttling or additional qubit-addressing optics, greatly simplifying the system architecture and allowing straightforward integration with existing trapped-ion quantum computers.
\end{abstract}

\maketitle

\section{Introduction}\label{sec:intro}
Applications of quantum information science, from computing and simulation~\cite{PRX_QComp,PRX_QSim} to communication~\cite{PRX_QComm} and sensing~\cite{PRL_QSense}, are predicated upon high-fidelity initialization and measurement of quantum systems~\cite{anHighPRL2022,loschnauerScalable2024,bluvsteinLogicalNature2024,smithSingleQubitPRL2025,sotirovaHighfidelity2024} such as qubits.
With increasing demand for interaction complexity and fidelity, classical feedback has become a critical tool for quantum control, with prime examples of auxiliary quantum sensing~\cite{Majumder2020} and quantum error correction~\cite{Roffe2019}.
This approach usually requires measurements on a subsystem of auxiliary qubits while not affecting the coherence of the remaining data qubits.

In atomic systems, mid-circuit measurement and reset (MCMR) operations are performed by scattering light from laser beams and detecting the resulting fluorescence.
These dissipative operations typically drive strong atomic transitions with a relatively large spectral bandwidth,
where even a single photon from the driving laser or re-radiated light can destroy quantum information stored in nearby data qubits.
Isolation can be achieved by using multiple atomic species~\cite{Schmidt2005,negnevitskyRepeatedNature2018,pino2021, Singh2023},
shuttling atoms in and out of measurement/reset zones~\cite{zhuInteractiveNat.Phys.2023, mosesRaceTrackPhys.Rev.X2023, Bluvstein2022},
or spatially discriminating auxiliary and data atoms with well-focused laser beams~\cite{motlakuntaPreservingNatCommun2024}.
However, these methods increase the system complexity and can significantly degrade the speed of operations~\cite{mosesRaceTrackPhys.Rev.X2023}.

\begin{figure}
  \includegraphics[width=0.435\textwidth]{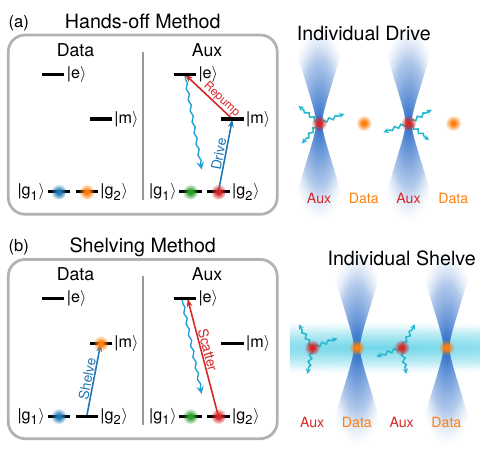}
  \caption{
    MCMR sequences via metastable intermediate states. The energy levels here are abstract and may be applicable to different ion types. Most of the qubit operations in circuits are done on the two ground states $|g_1\rangle$ and $|g_2\rangle$. The metastable $|m\rangle$ state has a long enough lifetime such that its decay can be ignored during the sequence. Finally, a short-lived excited state $|e\rangle$ can be used for photon scattering. Depending on the type of the sequence, different $|e\rangle$ may be chosen for either a cycling or non-cycling transition. (a) The hands-off method does not disturb the data qubit and drives the auxiliary ion to $|m\rangle$ as an intermediate state for photon scattering on $|e\rangle$. (b) The shelving method hides the data qubit in state $|m\rangle$ and measures the auxiliary qubit.
    \label{fig:mcmr-methods}
  }
\end{figure}

\begin{figure*}
  \includegraphics[width=0.784\textwidth]{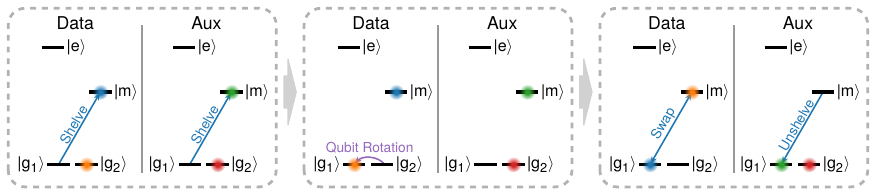}
  \caption{
    Example individual metastable state control sequence implemented with qubit rotation.
    This sequence achieves the same final state as the one in~Fig.~\ref{fig:mcmr-methods}b
    while only requiring individual qubit gates and global shelving operations.
    \label{fig:qubit-rotation}
  }
\end{figure*}

In this paper, we demonstrate MCMR operations in an existing quantum computing system without increasing its hardware complexity.
We propose general MCMR sequences using two different types of qubits in the same atomic species,
in a version of the so-called \textit{omg} qubit architecture~\cite{allcockOmgAppliedPhysicsLetters2021,lisMidcircuitPhys.Rev.X2023,maHighfidelityNature2023,grahamMidcircuitPRX2023}.
These MCMR sequences allow MCMR operations to be controlled from a high-level software layer
and without shuttling, specific spatial ordering of the atoms,
or additional individual qubit addressing optical elements.

In sections~\ref{sec:theory:mcmr} and~\ref{sec:theory:metastable-ind}, we discuss MCMR methods implementation and metastable state control to achieve selectivity in accessing the metastable state in long ion chains.
We experimentally demonstrate these techniques on our system in section~\ref{sec:experiment}
and conclude with a discussion of the techniques and outlook in section~\ref{sec:discussion}.

\section{MCMR using metastable states}\label{sec:theory:mcmr}
We implement MCMR operations using two different approaches.
The \textbf{\emph{hands-off method}}~\cite{lindenfelserCoolingNJP2017,chenNoninvasive2025} does not disturb the data qubits and drives the auxiliary atoms to a metastable qubit state and then cycles them back to the ground state via a short-lived excited state~(Fig.~\ref{fig:mcmr-methods}a).
The \textbf{\emph{shelving method}}~\cite{manovitzPRXQ2022} shelves the data qubit to metastable states that are isolated from photon scattering, thus protecting the quantum information~(Fig.~\ref{fig:mcmr-methods}b).

Although both methods can achieve high fidelity MCMR, they have different error and speed characteristics.
The hands-off method performs all MCMR operations on the auxiliary qubit, thereby limiting exposure of the data qubits to shelving errors.
However, this isolation relies on low coupling between the auxiliary drive and the data qubits throughout the measurement or reset sequence.
In addition, measurement may be slower, limited by the speed of the auxiliary qubit metastable transition.
The shelving method hides the data qubit from the MCMR operations and can provide better crosstalk protection during measurement and reset, but requires high fidelity shelving and high coherence of the metastable state.
A comparison between the two methods is therefore a trade-off between the coherence of the shelved state and the unwanted coupling of the metastable state drive and may even depend on whether the subsystem is being measured or just reset, as the measurement requires more photon scattering.

\section{Individual metastable states access}\label{sec:theory:metastable-ind}
The use of additional metastable states better isolates the data qubits during MCMR operations but requires individual control to drive specific qubits to the metastable state.
This can be accomplished with individual addressing beams already in place for quantum gate operations without adding complexity to the setup.
One implementation of such control exploits the gate beams to spectrally shift (dress) target qubits (\textit{dressing implementation}).
A simple global beam can then be used to selectively drive the ions to the metastable state,
by tuning to either the shifted or the bare resonance~\cite{lisMidcircuitPhys.Rev.X2023,jamiedleppardEnablingBull.Am.Phys.Soc.2024,norciaMidcircuitPRX2023,huSiteSelectivePRL2025}.
Alternatively, the combination of the qubit states and the metastable states can be treated as a qudit system~\cite{lowControlnpjQuantumInf2025,nikolaevaScalable2025},
and individual metastable state control can be viewed as a special case of a single-qudit operations.
Since arbitrary individual qudit gates can be achieved by combining global $\pi$-pulses and individual control on two of the qudit levels (\textit{qubit rotation implementation}),
individual drive to the metastable state can be implemented with a sequence of single qubit gates and global drive to the metastable state.
The optimal compiled sequence depends on the exact qudit operations to be implemented,
and an example of such a composite sequence to shelve only one ion is shown in~Fig.~\ref{fig:qubit-rotation}.

Comparing the two implementations of individual metastable state control,
the dressing can be applied to both hands-off or shelving MCMR methods.
In both cases, the auxiliary qubit should be dressed for better data qubit protection.
The qubit rotation implementation, on the other hand, usually involves driving all qubits to the metastable states.
This unconditionally perturbs the data qubits and is therefore not appropriate for the hands-off method.
However, since the only required individual operations are single qubit gates,
this implementation is more generally applicable
and may benefit from optimized gates for coherent quantum circuits.


\section{Experimental implementation}\label{sec:experiment}
\begin{figure}
  \includegraphics[width=0.45\textwidth]{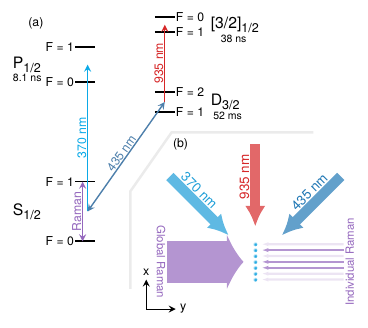}
  \caption{
    Overview of the experiment.
    (a) Relevant energy levels and supported operations on the $^{171}\mathrm{Yb}^+$ ions. The normal reset and measurement operation is done on the $370\ \mathrm{nm}$ transition between $S_{1/2}$ and $P_{1/2}$ with population leakage into the $D_{3/2}$ repumped back with the $935\ \mathrm{nm}$ transition via the $[3/2]_{1/2}$ state. The metastable $D_{3/2}$ state can also be directly populated using the $435\ \mathrm{nm}$ transition.
    (b) Beam geometry relative to the ion chain. The $370\ \mathrm{nm}$ and $435\ \mathrm{nm}$ beams are sent in $45^\circ$ relative to the ion chain parallel to the trap surface and the $935\ \mathrm{nm}$ beam is sent in along the chain. All of these are global beams that illuminate the whole chain. The Raman beam path consists of a global beam perpendicular to the chain and an array of counter-propagating individual beams to target each ions separately.
    \label{fig:setup}
  }
\end{figure}
Here, we demonstrate these capabilities experimentally using a fully reconfigurable trapped-ion quantum computing system~\cite{monroeDemonstrationPRL1995,kozhanovNextgenerationOpt.Quantum20Conf.Exhib.2023Pap.QM3A22023}.
A chain of up to 32 \Yb ions can be trapped using a Phoenix surface trap fabricated by Sandia National Laboratory~\cite{revellePhoenixArXiV2020,kozhanovNextgenerationOpt.Quantum20Conf.Exhib.2023Pap.QM3A22023}.
For all the experimental demonstrations,
we use two ions separated by $4.5\ \mathrm{\mu m}$,
which is the same as the distance used for gate operations in a longer chain.

As shown in the level diagram of~Fig.~\ref{fig:setup}a,
we use the magnetic field insensitive qubit $|1\rangle\equiv|F=1,m_F=0\rangle$~and~$|0\rangle\equiv|F=0,m_F=0\rangle$ in the ground $S_{1/2}$ manifold.
The qubit is coherently controlled using $355\ \mathrm{nm}$ beams~\cite{Cetina2022} - one global beam addressing all ions in the chain and an array of counter-propagating,
tightly-focused beams addressing each ion individually~\cite{nagerlLaserPRA1999,Cetina2022}~(Fig.~\ref{fig:setup}b).
All other beams used in the experiment are global beams illuminating the entire chain and are aligned parallel to the trap surface~(Fig.~\ref{fig:setup}b).
Dissipative operations (Doppler cooling as well as the global reset and measurement operations of the qubit states) are done using the $370\ \mathrm{nm}$ transition with a state preparation and measurement~(SPAM) error of $0.2\ \%$ ($0.5\ \%$) for the $|0\rangle$ ($|1\rangle$) states.

\subsection{Individual metastable states access}\label{sec:experiment:metastable-ind}
\begin{figure}
  \includegraphics[width=0.4\textwidth]{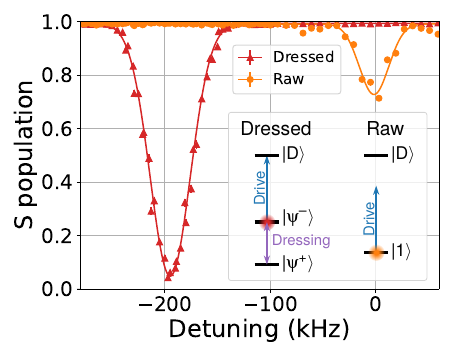}
  \caption{
    Simultaneous $D_{3/2}$ spectrum on two ions with individual dressing on one ion.
    Inset: illustration of the measurement protocol.
    \label{fig:dstate-spec}
  }
\end{figure}
For the metastable state used to implement MCMR,
we select the $|D \rangle \equiv | D_{3/2},F=2,m_F=0\rangle$ state
coherently driven to from the ground state by a $435\ \mathrm{nm}$ beam~(Fig.~\ref{fig:setup}b)
with a $T_2$ coherence time of $10\ \mathrm{ms}$.

Individual driving to the $|D\rangle$ state is achieved using the qubit-rotation or the dressing implementations~(Sec.~\ref{sec:theory:metastable-ind}).
Since the qubit-rotation implementation of individual metatable state control
requires only single qubit operations,
its distinction of the ions is limited by our single qubit gate fidelity at $>99\ \%$.
On the other hand, due to the minimum scalar Stark shift produced by our Raman beams,
we use a near-resonant Raman transition to implement dressing of ions~(See Appendix~\ref{appendix:raman-dressing}).
We demonstrate this by measuring the $D$ state spectrum on two ions simultaneously.
One ion is prepared in the a dressed state,
whereas the other ion is not dressed and remains in $|1\rangle$~(Fig.~\ref{fig:dstate-spec}).
We drive the ions to the $D$ state using a Blackman time profile
to suppress off-resonant excitation~\cite{blackman1958}.
The pulse length is chosen to drive a $\pi$ rotation for the dressed state
and is the same for both measurements to allow easier comparison.
Since the bare state is more strongly coupled than the dressed one,
this pulse length overdrives the bare resonance,
which appears as a smaller peak as a result of the Blackman pulse shape.

\subsection{Shelving method}\label{sec:experiment:shelving}
\begin{figure}
  \includegraphics[width=0.5\textwidth]{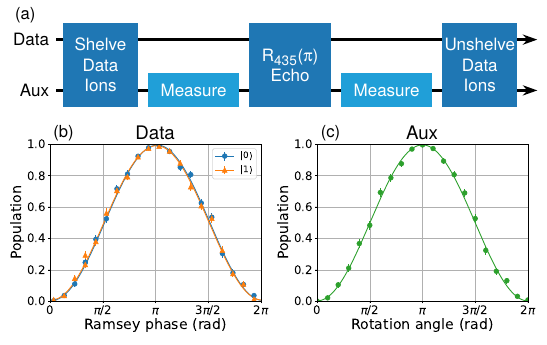}
  \caption{
    Mid-circuit measurement with shelving method~(Fig.~\ref{fig:mcmr-methods}b) to the $D_{3/2}$ manifold. (a) Experimental sequence. First, the data ions are shelved to the $|D\rangle$ state. The auxiliary ion is measured using the $370\ \mathrm{nm}$ transition with a pause for the global spin echo to improve data ion fidelity. After the measurement, the data ions are unshelved back to the ground states. The shelving and unshelving of the data ions may use either the dressing implementation or a simplified qubit rotation sequence.
    (b) Ramsey phase scan demonstrating that the mid-circuit measurement of the auxiliary ion using the qubit rotation implementation preserves the coherence of the data ion coherence. The contrast of the Ramsey fringe shows a data ion fidelity of $98.8(6)\ \%$ ($98.2(7)\ \%$) when the auxiliary ion was in the $|0\rangle$ ($|1\rangle$) state.
    (c) Mid-circuit measurement on the auxiliary ion which is initialized in a superposition of $|0\rangle$ and $|1\rangle$ with a single qubit rotation $R_x(\theta)$ with variable rotation angle $\theta$. The result shows a measurement fidelity of $99.6(2)\ \%$ which is identical to that of a normal measurement.
    \label{fig:detect}
  }
\end{figure}
We demonstrate the shelving method~(Fig.~\ref{fig:detect}a) by implementing mid-circuit measurement on an ion chain that is cooled to the motional ground state on all three axes.
Since the destination of the auxiliary $|0\rangle$ state does not affect the measurement result as long as it is not in $|1\rangle$,
the shelving sequence can be significantly simplified
by allowing the $|0\rangle$ state for all the ions to be shelved globally.
Individual control during the shelving step is therefore mainly needed
to differentiate the $|1\rangle$ state between the two ion types,
which can be realized with both the dressing and the qubit rotation implementations.

In the case of the dressing implementation, the shelving sequence first dresses the auxiliary ions with the Raman beams and drives only the data ion's $|0\rangle$ state to $|D\rangle$.
The dressing is done with a detuning of $\delta=0.1\Omega$ and uses the corresponding sequence in Table~\ref{tab:dress-rot}~(Appendix~\ref{appendix:dress-rotation}) for rotating in and out of the dressed states.
After this, the data ion's $|1\rangle$ state is driven to $|0\rangle$ using our typical Raman transition.
The final mapping is: $|0\rangle\rightarrow|D\rangle$, $|1\rangle\rightarrow|0\rangle$ on the data ions, and no operation on the auxiliary ions.

For the qubit rotation implementation, the shelving sequence uses a global drive from $|0\rangle$ to $|D\rangle$
followed by the same Raman step to transfer the data ion's $|1\rangle$ to $|0\rangle$.
The mapping is: $|0\rangle\rightarrow|D\rangle$, $|1\rangle\rightarrow|0\rangle$ on the data ions,
and $|0\rangle\rightarrow|D\rangle$ on the auxiliary ions.
Note that with this simplification, the dressing sequence is strictly more complex than the qubit rotation one
since it requires additional Raman dressing during the drive to the $|D\rangle$ state.
However, this simplification relies on the $F=1$ and $F=0$ ground state hyperfine structure,
which allows the data qubit to remain in $|0\rangle$ during measurement,
and does not apply to all ions, e.g. $^{137}\mathrm{Ba}^+$,
where the dressing implementation may be preferred.

After the shelving step, the detection of the auxiliary ions then proceeds to the $370\ \mathrm{nm}$ transition. We use a $\pi$ pulse between $|0\rangle$ and $|D\rangle$ in the middle of the detection time to perform spin echo on the data ion. This echo is done with the global $435\ \mathrm{nm}$ beam, but it does not affect the measurement result on the auxiliary ion, since it is not coupled to the $|1\rangle$ state. Finally, we apply the inverse of the shelving sequence to bring the data ion back to the $|0\rangle$ and $|1\rangle$ states. The shelving and unshelving process each takes $26\ \mathrm{\mu s}$ ($47\ \mathrm{\mu s}$) for the qubit rotation (dressing) implementation, and the photon scattering step takes $140\ \mathrm{\mu s}$.

We characterize the fidelity of the mid-circuit measurement for the auxiliary ion in both the $|0\rangle$ and $|1\rangle$ states by performing a single qubit rotation on the auxiliary ion before the entire measurement sequence is applied~(Fig.~\ref{fig:detect}c). We show a measurement fidelity of the $|0\rangle$ ($|1\rangle$) state of $99.7(1)\ \%$ ($99.5(3)\ \%$) regardless of the shelving implementation. This result is indistinguishable from the end-of-circuit measurement fidelity in our setup.
In addition, the effect of the auxiliary ion measurement sequence on the data ion was measured via a Ramsey experiment. For the qubit rotation implementation, the data ion fidelity extracted from the Ramsey fringe in Fig.~\ref{fig:detect}b is $98.8(6)\ \%$ ($98.2(7)\ \%$) when the auxiliary ion was in $|0\rangle$ ($|1\rangle$) state. The fidelity is limited by the coherence of the $S_{1/2} \rightarrow D_{3/2}$ transition. For the dressing implementation, the data qubit fidelity is $91.7(4)\ \%$ when the auxiliary ion is in either $|0\rangle$ or $|1\rangle$.
The low fidelity of this implementation is due to crosstalk from the Raman dressing beams, as expected. It could be improved by changing the dressing parameters, e.g. increasing the dressing detuning. However, we do not optimize this method further due to its intrinsic disadvantage on our system compared to the qubit rotation implementation. The fidelities of the data qubit for both implementations are SPAM corrected.

\subsection{Hands-off method}\label{sec:experiment:hands-off}
\begin{figure}
  \includegraphics[width=0.5\textwidth]{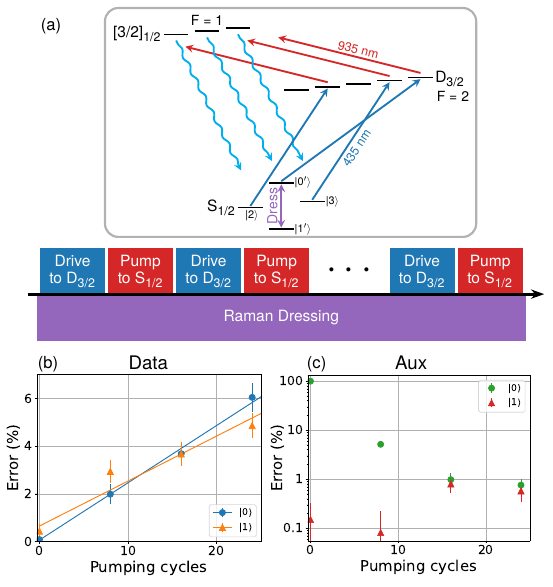}
  \caption{
    Mid-circuit reset with hands-off method~(Fig.~\ref{fig:mcmr-methods}a) using the $D_{3/2}$ manifold. (a) Experimental sequence. The auxiliary ion is dressed with the Raman beams during the whole sequence. The population in the undesired states ($|0'\rangle$, $|2\rangle$ and $|3\rangle$) are driven to the $D_{3/2}$ states by $435$~nm light and then pumped back down to the $S_{1/2}$ states by $935$~nm. This process repeats until all the population is pumped to the $|1'\rangle$.
    (b, c) Infidelity of the data ion and the auxiliary ion after pumping, demonstrating a $0.9(3)\ \%$ error on auxiliary ion and $3.7(5)\ \%$ on data ion after $16$ cycles of pumping.
    \label{fig:pump}
  }
\end{figure}
We demonstrate the hands-off method by performing mid-circuit reset. The dressing is done with $\delta/\Omega=0.5$. The corresponding sequence in Table~\ref{tab:dress-rot}~(Appendix~\ref{appendix:dress-rotation}) is used to rotate in and out of the dressed states.
In order to fully reset the ion to a single state, we also need to pump $|2\rangle\equiv|F=1,m_F=-1\rangle$ and $|3\rangle\equiv|F=1,m_F=1\rangle$ that are outside the qubit space.
We drive these states to the $|D_{3/2}, F=2, m_F=-1\rangle$ and $|D_{3/2}, F=2, m_F=1\rangle$ respectively using a $435$~nm global beam. These transitions are isolated by a sufficient Zeeman shift ($5\ \mathrm{MHz}$) from any transitions on the states populated by the data ion~(Fig.~\ref{fig:pump}a). After driving the desired auxiliary ion states to $D_{3/2}$, we use $935\ \mathrm{nm}$ light to pump $D_{3/2}$ states back to $S_{1/2}$ to complete a pumping cycle. We perform the reset with different numbers of pumping cycles. Fig.~\ref{fig:pump}c shows the pumping error measured on the auxiliary ion starting with either the $|0\rangle$ or $|1\rangle$ initial states. 16 pumping cycles is sufficient to complete the auxiliary qubit reset operation with a $0.9(3)\ \%$ error. Fig.~\ref{fig:pump}b shows the error on the data ion during the pumping process with $3.7(5)\ \%$ error after 16 cycles. Based on a master equation simulation, the infidelity is mainly caused by off-resonant coupling to the $D_{3/2}$ state and the motional excitation of the ions. Due to technical limitations, the ions are not cooled to the ground motional state for this reset demonstration. The fidelity on both the data and auxiliary ions are SPAM corrected.

\section{Discussion}\label{sec:discussion}
We studied how \textit{omg} architecture can be used for MCMR in a trapped-ion quantum computing system with long chains of ions.
We experimentally demonstrated both the shelving and hands-off methods for MCMR
alone with the different ways to drive individual ion to metastable state
on our $^{171}\mathrm{Yb}^+$ quantum computing system.
The fidelity achieved is limited mainly by technical sources
and can be straightforwardly improved with better laser locking,
laser power and polarization control,
and performing ground state cooling on the ion chain.
Switching to a metastable state with longer lifetime,
e.g. $F_{7/2}$ for $\mathrm{Yb}^+$ or $D_{3/2}$ and $D_{5/2}$ for $\mathrm{Ba}^+$,
could also improve the fidelity significantly.

While the photon scattering during measurement causes recoil heating,
it is negligible for reset, which only requires few photons to be scattered.
Furthermore, heating during measurement can be minimized
by integrating EIT~\cite{ImagingPRL2023}
or sideband cooling~\cite{SiteResolvedPRL2015} into the detection process.

Although all of the methods provide a speedup by avoiding the time-consuming shuttling
and re-cooling steps without the need for additional individually addressed beams,
the correct choice for a particular system depends on a number of factors.

From the atomic-physics side, the shelving method requires a coherent time
for the shelved qubit state that is significantly longer than the measurement/reset time
whereas the hands-off method could work well even with a short-lived metastable state
with sub-millisecond lifetime.
Moreover, the state used for shelving must also not participate in the photon scattering,
which is one of the reasons our mid-circuit reset is done using the hands-off method.

On the technical side, the shelving method incoperates fewer operations
that require individual addressing and can therefore be beneficial
if the fidelity is limited by addressing crosstalk.
The available Rabi frequencies for both the individual beams
as well as the global metastable transition beam may also affect the choice.
When dressing is used, the drive to the metastable state needs to be significantly slower
than the shift produced and can therefore be slower compared to qubit rotation
if the metastable state transition could otherwise achieve a higher Rabi frequency.
By extension, the hands-off method, which implies dressing,
could also be slower if the coupling to the metastable state cannot achieve a fast rate.
This is mainly a concern when used for detection
due to the larger number of photon scattering needed
and can be overcome with higher laser power.

Finally, although our scheme does not require any new individual addressing beam path,
some addition to the system is of course still necessary.
The most straightforward requirement is the new lasers wavelengths
to access the desired metastable states,
which limited our demonstration to the $D_{3/2}$ state in \Yb ions,
due to the absence of beams to couple to any other metastable states.
Moreover, using the hands-off method for measurement may also require
an additional global laser beam to scatter photons from the metastable state,
as well as the detection of the resulting photons.
This is one of the main reasons
that prevents us from using the hands-off method for measurement,
but it could be overcome in the future
either by detecting the $297\ \mathrm{nm}$ photons emitted
when pumping the ion back from $D_{3/2}$ with $935\ \mathrm{nm}$,
or adding a $2.438\ \mathrm{\mu m}$ laser
to pump the $D_{3/2}$ state back via the $P_{1/2}$ state,
allowing the detection of the resulting $370\ \mathrm{nm}$ photons \cite{kramidaNIST1999}.

\section{Acknowledgments}
This work is supported by the DARPA Measurement-based Quantum Information and Transduction program (HR0011-24-9-0357), the DOE Quantum Systems Accelerator (DE-FOA-0002253), the NSF STAQ Program (PHY-1818914), and the NSF Major Research Instrumentation program (2117530).

\appendix
\section{Metastable states in \Yb}\label{appendix:metastable}
The \Yb ions present multiple metastable states as candidates for the metastable states
used in our MCMR scheme, $D_{3/2}$, $D_{5/2}$ and $F_{7/2}$.
We use the metastable $D_{3/2}$ states for our initial demonstration
since it is easily accessible with a single photon transition
with a longer lifetime and cleaner decay path among the two $D$ states.
However, this state does participate in the normal photon cycling
on the $\mathrm{D1}$ line due to a $0.5\%$ branching ratio from the $P_{1/2}$ state
and this leakage is addressed using the $935\ \mathrm{nm}$ repumping beam.
This beam contains a fixed tone addressing the $D_{3/2}(F=1)$ hyperfine state,
which participates in all of the dissipative operations,
and a controllable sideband addressing the $D_{3/2}(F=2)$ hyperfine state
that is switched on during cooling and pumping and can be switched off during detection.

Since the $D_{3/2}$($F=1$) state is continuously driven
by the fixed $935\ \mathrm{nm}$ tone, only the $D_{3/2}$($F=2$) state is usable
for coherent operation during our MCMR scheme.
Depending on the lock point, we are able to drive either the $F=1$ or $F=0$ hyperfine levels in the $S_{1/2}$ manifold to the $D_{3/2}$ state.
Due to off-resonant scattering from the fixed $935\ \mathrm{nm}$ tone,
the lifetime of the $D_{3/2}$($F=2$) state is shortened to $15\ \mathrm{ms}$ from the $53\ \mathrm{ms}$ natural lifetime.
This contributes to the $10\ \mathrm{ms}$ $T_2$ coherence time.
These technical issues limit the fidelity of our MCMR sequence
and could be improved with better control of the $935\ \mathrm{nm}$ beam (by adding the ability of turning it off)
and better locking of the $435\ \mathrm{nm}$ laser which should result in reduced laser phase noise and improve the coherence time of the optical qubit.

\section{Near resonance Raman dressing}\label{appendix:raman-dressing}
The dressing implementation of individual metastable state control uses the individual gate control beams to dress the auxiliary ion and change the resonance of the $435\ \mathrm{nm}$ transition to the $|D\rangle$ state.
However, the wavelength of Raman beams used in our system is selected to minimize the scalar AC Stark shift on the $S_{1/2}$ state in order to improve gate fidelity caused by the differential Stark shift during gate operations.
Additionally, the Raman light is far detuned from any $D$ state transitions, and therefore we cannot produce a significant shift on the shelving transition by simply turning on the individual Raman beams.
Instead, we rely on dressing the ions with a Raman transition.

The Hamiltonian of an ion driven on a Raman transition with a Raman Rabi frequency of $\Omega$ and a detuning of $\delta$ can be written as:
\begin{align}
  H=&\delta|0\rangle\langle0|+\frac{\Omega}{2}\left(|0\rangle\langle1|+|1\rangle\langle0|\right)
\end{align}
This produces two dressed states
\begin{align}
  |\psi_\pm\rangle=&\frac{\sqrt{1\pm \delta/\Omega_g}|0\rangle\pm\sqrt{1\mp \delta/\Omega_g}|1\rangle}{\sqrt2}
\end{align}
with corresponding frequency shifts from the bare resonance,
\begin{align}
  \Delta_\pm=&\frac{\delta\pm\Omega_g}{2}
\end{align}
where $\Omega_g\equiv\sqrt{\Omega^2+\delta^2}$ is the generalized Rabi frequency.
The two dressed states are shifted from the bare resonance with opposite signs,
and the absolute values of the shifts are different when $\delta\neq0$.
The one with a smaller shift is usually called the Stark shifted state
whereas the resonance with a larger shift corresponds to a multi-photon transition.
This asymmetry gives rise to different requirements for the dressing beam
when it is used in the hands-off or shelving methods.
For the hands-off method, where the dressed auxiliary ions are driven to the metastable state,
only the dressed state needs to be sufficiently shifted to avoid driving the data ions,
this allows a larger shift asymmetry, i.e. a large $\delta$, to be used.
In contrast, the shelving method drives the data ions to the metastable state
while leaving both dressed states untouched for the auxiliary ions,
therefore favoring a more symmetric shift and a smaller $\delta$.
Moreover, in order to reduce the effect of crosstalk from the dressing beam on the data ions,
$\delta$ should be significantly larger than the crosstalk Raman Rabi frequency.

\section{Robust switching to the dressed basis}\label{appendix:dress-rotation}
The process of mapping the two qubit states to and from the two dressed states accurately
and robustly on the auxiliary ion is important for achieving a high MCMR fidelity.
In the Stark shift regime, i.e. far-detuned dressing,
this is typically done by turning on the dressing beam adiabatically.
However, when the detuning is comparable to or smaller than the Rabi frequency,
the time it takes to achieve the desired level of adiabaticity becomes impractical.
\begin{table}
  \begin{NiceTabular}{|c|c||c|c|}
    \hline
    \multicolumn{2}{|c||}{}&Sequence 1&Sequence 2\\\hhline{====}
    Target&$\delta/\Omega$&$0.1$&$0.5$\\\hline
    \multirow{3}{*}{Pulse 1}&$\delta_1/\Omega$&$0.066045984$&$0.865404511$\\\cline{2-4}
    &$\tau_1\Omega$&$6.101$&$2.482$\\\cline{2-4}
    &$\phi_1$&$4.697118972032$&$5.922001257829$\\\hline
    \multirow{3}{*}{Pulse 2}&$\delta_2/\Omega$&$1.738387469$&$-0.467035375$\\\cline{2-4}
    &$\tau_2\Omega$&$3.372$&$5.03$\\\cline{2-4}
    &$\phi_2$&$0.0$&$3.710143207203$\\\hline
    \multirow{3}{*}{Pulse 3}&$\delta_3/\Omega$&$-0.304962921$&$0.087101714$\\\cline{2-4}
    &$\tau_3\Omega$&$7.47$&$4.478$\\\cline{2-4}
    &$\phi_3$&$2.106261199596$&$0.56914785185$\\\hline
    \multirow{2}{*}{Errors}&$E_0$&$2.5\times10^{-4}$&$2.9\times10^{-7}$\\\cline{2-4}
    &$E_1$&$1.9\times10^{-10}$&$2.3\times10^{-7}$\\\hline
  \end{NiceTabular}
  \caption{
    Rotation sequence parameters and performance. Each of the composite pulse sequence contains three Raman pulses with detuning $\delta_i$, time $\tau_i$ and $\phi_i$ ($i=1,2,3$). The detunings and times are recorded in the table as unitless number relative to the nominal Rabi frequency $\Omega$. $E_0$ is the total error in the Pauli $X$ and $Y$ coefficient for a crosstalk Rabi frequency of $0$ to $5\ \%$ of the nominal Rabi frequency, and $E_1$ is the error for a fluctuating target ion Rabi frequency from $90\ \%$ to $105\ \%$ of nominal. The significant digits recorded in the table reflects the hardware resolution used in our experiment.
    \label{tab:dress-rot}
  }
\end{table}

\begin{figure}
  \includegraphics[width=0.45\textwidth]{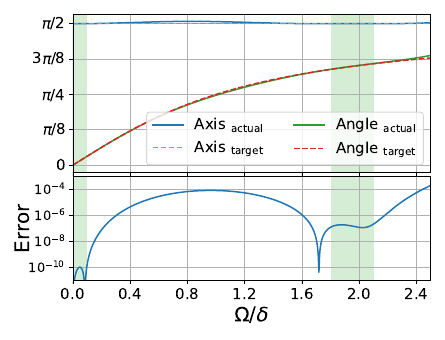}
  \caption{
    Performance of an example rotation sequence between the undressed and dressed state for $\delta/\Omega=0.5$. The rotation axis, angles (upper) and the error of the rotation (lower) is plotted as a function of the Rabi frequency. The green band marks the relevant range of Rabi frequency.
    \label{fig:dress-rotate}
  }
\end{figure}

Instead, we use a single qubit rotation to map between the two bases.
Since the target unitary to rotate into the dressed basis
depends on the power of the Raman beams,
$U(\Omega) = |\psi^+\rangle\langle0|+|\psi^-\rangle\langle1|+h.c.$,
we use numerical optimization to construct a composite pulse
that produces the desired unitary robustly for a range of Rabi frequencies
around the nominal value.
Additionally, for neighboring ions that may experience a low Rabi frequency
due to Raman beam crosstalk,
the sequence will also rotate them into the corresponding dressed states.
This ensures that crosstalk errors would be predominantly a phase error
and can be corrected with spin echo pulses during the MCMR sequence.
After the dressing step is completed,
time reversal of the same sequence can then be used to map the qubit back
to the original basis with the same robustness property.
Table~\ref{tab:dress-rot} lists the sequence parameters designed
for different nominal detuning values and their performance.
As an example, Fig.~\ref{fig:dress-rotate} plots the calculated rotation
performed by the pulse sequence for $\delta/\Omega=0.5$
and the error it produces from the expected one,
showing an expected error of no more than $10^{-6}$
in the relevant range of Rabi frequencies.

\bibliographystyle{apsrev4-2}
\bibliography{library.bib}

\end{document}